\documentstyle[psfig]{elsart}

\begin{document}

\begin{frontmatter}

\title{Silicon Pad Detectors for the PHOBOS Experiment at RHIC}

\author[UIC]{R. Nouicer},
\author[ANL]{B. B. Back},
\author[UIC]{R. R. Betts},
\author[MIT]{K. H. Gulbrandsen},
\author[UIC]{B. Holzman},
\author[UIC]{W. Kucewicz},
\author[NCU]{W. T. Lin},
\author[MIT]{J. M\"{u}lmenst\"{a}dt},
\author[MIT]{G. J. van Nieuwenhuizen},
\author[MIT]{H. Pernegger},
\author[UIC]{M. Reuter},
\author[MIT]{P. Sarin},
\author[MIT]{G.S.F. Stephans},
\author[MIR]{V. Tsay},
\author[MIT]{C. M. Vale},
\author[MIT]{B. Wadsworth},
\author[ANL]{A. H. Wuosmaa},
\author[MIT]{B. Wyslouch}

\address[UIC]{ Department of Physics, M/C 273,
		University of Illinois at Chicago,
		845 West Taylor St.,
		Chicago, IL 60607-7059 , USA}
\address[ANL]{Physics Division,
		Argonne National Laboratory,
		9700 South Cass Ave.,
		Argonne, IL 60439-4843, USA}
\address[MIT]{Massachusetts Institute of Technology,
                77 Mass. Ave.,
                Cambridge, MA 02139, USA}

\address[NCU]{High Energy Physics Group,
		National Central University,
		Department of Physics,
		32054 Chung-Li, Taiwan}
\address[MIR]{Miracle Technology Co. Ltd.,
		Hsin-Chu, Taiwan}

\begin{abstract}
The PHOBOS experiment is well positioned to obtain crucial information about
relativistic heavy ion collisions at RHIC, combining 
a multiplicity counter with a multi-particle spectrometer.
The multiplicity arrays will measure the charged particle multiplicity over the full solid angle. 
The spectrometer will be able to identify particles at mid-rapidity.
The experiment is constructed almost exclusively of silicon pad detectors. 
Detectors of nine different types are configured in the multiplicity and vertex 
detector (22,000 channels) and two multi-particle spectrometers
(120,000 channels). 
The overall layout of the experiment, testing of the silicon sensors 
and the performance of the
detectors during the engineering run at RHIC in 1999 are discussed. 
\end{abstract}
\end{frontmatter}

\section{Introduction}
\vspace*{-0.4cm}
The Relativistic Heavy Ion Collider (RHIC) at Brookhaven
National Laboratory will open a new horizon for Ultra-Relativistic
Heavy-Ion physics, to explore highly excited dense nuclear matter
under controlled laboratory conditions at a center-of-mass
energy of 200 AGeV. At this energy,
it is speculated that a plasma of deconfined quarks and gluons (QGP) will be
formed following the initial series of nucleon-nucleon collisions. 
The QGP subsequently expands, cools and passes into the normal
hadronic phase which itself expands until the hadrons cease to
interact with each other at the freeze-out stage.
A number of signatures of the formation of the deconfined phase
have been proposed \cite{singh,harris,alam}.
It is generally thought that a single signature will be insufficient
to provide the evidence for the QGP. Rather, the simultaneous
observation of several of the proposed signatures, particularly
on an event-by-event basis, will be required.
PHOBOS seeks to address these questions in a timely
fashion by focusing on measurements of hadronic observables for a very large
sample of events.
\vspace*{-0.5cm}
\section{PHOBOS detector}
\vspace*{-0.2cm}
The PHOBOS detector consists
of a 4$\pi$ multiplicity array, vertex finding detectors,
two multi-particle tracking spectrometers, a set of plastic
scintillator time-of-flight (TOF) walls, and trigger detectors.
A schematic diagram of the experiment, showing most of 
these elements is shown in Fig. \ref{fig:phobos}.
The experiment is constructed almost exclusively of silicon pad detectors.
Which have the advantage as providing simultaneously good position resolution, 
low multiple scattering in the detector and good energy loss resolution.
The multiplicity array consists of an octagonal barrel
of silicon pad detectors surrounding the beam pipe, and six rings
of silicon pad detectors. This array covers
the pseudo-rapidity range of $\vert \eta \vert$ $\le$ 5.3. 
The multiplicity detector will provide
event-by-event charged-particle multiplicity distributions
which can be used to select interesting events for further study.
The multiplicity
distributions are interesting in their own right, containing information
on multiplicity fluctuations and correlations, which potentially
can be related to some of the proposed signatures of the QGP.
The vertex detectors consist of two sets of
silicon pad sensors situated above and below the beam line in the
interaction region. Each set consists of two layers 
of silicon sensors with 4 sensors in the inner layer and 8 sensors in the 
outer layer as seen from the interaction region.
The silicon vertex finder will be able to determine the 
position of the interaction point with an accuracy of 50~$\mu$m. 
 The spectrometers consist of planes silicon pad detectors
positioned on either side of the beam. Some of these planes lie within a 2T magnetic field.
The spectrometers are able to measure and identify particles with
transverse momenta as small as to 50~MeV/c for pions.
The spectrometer will determine the particle momentum by measuring the 
track curvature and the particle type by the dE/dx method. 
The particle identification capability is further enhanced by two TOF 
walls behind the spectrometer.
Finally, trigger counters
consisting of two disks of 16 $\check{\rm C}$erenkov radiators, and 
two sets of plastic scintillator counters (paddle counters), are arranged around the beam pipe. 
The paddle counters are designed to trigger on peripheral collisions and to give
a first approximation of the event centrality.
\vspace*{-0.5cm}
\subsubsection{ Design of Silicon Pad Detectors}
\vspace*{-0.2cm}
The silicon pad detector is a single sided, AC coupled, detector that uses a double-metal 
layer to route the signals from the pads to the bonding area at the edge 
of the sensor. 
The wafer layout is conceptually similar to one used in CERN experiment WA98\cite{wilis}. 
The p+ implants are rectangular pads providing two-dimensional 
position information.
The implants are biased via polysilicon resistors 
which are connected to a common bias line. 
A schematic diagram of the cross section of one pad of the silicon 
detector is presented in Fig.  \ref{fig:structure}.  
The coupling capacitor between the first metal layer (Al) and the implant layer 
is formed by  a 0.2 $\mu$m thick layer of silicon Oxide-Nitride-Oxide (ONO). 
Each pad from this array is read out by a second metal 
line that connects to the Metal 1 
pad through a via and runs to the bonding pad.
The active area of the silicon wafer is surrounded by a guard ring. 
The double metal layer has the advantage that the readout structure is integrated 
on the sensor without additional material. The full detector surface can be used 
for routing the signal lines to the 
sensor edge to a single row or multiple rows of bonding pads,  permitting  
the use of standard readout chips. Nine different sensor types are used. 
The geometric characteristics of the different types of the sensors 
are presented in Table \ref{tab:physspecs}. 
A schematic layout of some of the silicon sensors used are shown 
in Fig. \ref{fig:octag} and more details about silicon sensors are presented in 
reference \cite{gerrit,heinz}.

\begin{table*}[hbt]
\caption{Physical specifications of the PHOBOS silicon pad
sensors}
\label{tab:physspecs}
\begin{tabular*}{\textwidth}{lllll}
\hline
Detector & Sensor 	& Active          & Number & Pad             \\
System   & Type   	& Area            & of     & Size            \\
         &      	& [mm${\rm{^2}}$] & Pads   & [mm$^{\rm{2}}$] \\
\hline
\hline
Spectrometer & 1       & 70.0 x 22.0 & 70 x 22 & 1.0 x 1.0 \\
             & 2       & 42.7 x 30.0 & 100 x 5 & 0.4 x 6.0 \\
             & 3       & 42.7 x 60.0 & 64 x 8  & 0.7 x 7.5 \\
             & 4       & 42.7 x 60.0 & 64 x 4  & 0.7 x 15.0 \\
             & 5       & 42.7 x 76.0 & 64 x 4  & 0.7 x 19.0 \\
\hline
Multiplicity & Octagon & 34.9 x 81.3 & 30 x 4 & 2.75 x 8.75  \\
             & Ring    & \ \ \ \ 3600  &  \ 8 x 8  & \ \ 20 - 105 \\
\hline
Vertex       & Inner   & 60.6 x 48.18 & 4 x 256 & 0.5 x 12.1 \\
             & Outer   & 60.6 x 48.18 & 2 x 256 & 0.5 x 24.1 \\
\hline
\end{tabular*}
\end{table*}

\vspace*{-0.2cm}
\subsubsection{ Testing and Acceptance Criteria of the Silicon Sensors}
\vspace*{-0.6cm}
The silicon sensors were produced by the Miracle Technology Co, 
Taiwan under supervision of the National Central University (NCU), Taiwan. 
The dicing and initial inspection was carried out at NCU. 
Subsequent testing of the characteristics of the Silicon wafers was conducted at 
the University of Illinois at Chicago (vertex, octagon and ring wafers) and at
Massachusetts Institute of Technology (spectrometer wafers). 
\\
This testing procedure was  performed using a computer controlled probe station 
running LabView (National Instruments), which can write the results directly into 
a central Oracle database \cite{data}. The probe station is equipped with Keithley 
units for current and voltage measurements as well as with capacitance-meters for 
capacitance measurements of the sensor. 
The following  tests are performed on each sensor :\\ 
{ \it - IV Test :}  The leakage current s measured as a function bias voltage applied to the  
active area and the guard ring separtely. 
The backplane voltage is applied through the vacuum jig. 
Sensors with less than 5 $\mu$A current for the full active area at full 
depletion voltage, typically around 70 V and with stable guard ring current are accepted.\\ 
{ \it  - PN Test :}  The sensor depletion voltage is measured on a p-n diode test structure 
adjacent to the sensor on the same production wafer.  This measured value for the 
depletion voltage was cross-checked with the results of tests with radioactive source 
carried out on the actual sensor. 
For most sensors, the depletion voltage was approximately 70 Volts.
\\
{\it - Polysilicon Test :}  The polysilicon bias resistor was measured for each wafer on two 
identical test structures on the wafer. These test structures consist of a chain of resistors 
identical to those used to bias each pad. We measure the current-voltage characteristic 
for both test structures 
and accept sensors with pad bias resistance larger than 1 M$\Omega$. 
 \\
{\it - Pinhole Test : } Pinholes, or connections between the two metal layers through the first 
dielectric layer, are detected by measuring the current 
between the bonding pad and the bias line. 
The bias line is kept at ground potential and the bond pads are held at - 5V. 
For functional oxides, the measured current is zero. 
This measurement was done for 
each pad.\\
{\it - CV Test : } The functionality of the readout line is checked by 
measuring the capacitance between each pad and the detector backplane, 
while the sensor is fully depleted. For functional lines a 
characteristic capacitance pattern is observed, which 
largely depends on the readout line geometry and routing. 
Broken or shorted lines clearly exhibit a 
different capacitance which can easily be distinguished from that of functional lines.
We allow a total of 3~$\%$ defective channels (sum of thin oxide defects and 
readout line defects) for each sensor.
\vspace*{-0.2cm}
\subsubsection{ Readout Electronics}
\vspace*{-0.2cm}
The readout of all PHOBOS silicon detectors is based on the VA-HDR-1 chip, a "Viking"
type chip manufactured by IDEAS \cite{ideas}. We use it in a 64-channel and a 128-channel 
version depending on the sensor granularity.
The VA-HDR-1 chip was chosen as it is commercially available and provides a large dynamic 
range, with input signals up to 400 fC, i.e. more than 100 times 
the energy loss of minimum ionizing particle (MIPs). 
The VA-HDR-1 chips are mounted on a hybrid. These hybrids are ceramic in the case of the vertex and 
spectrometer detectors, and multilayer printed circuit boards for the multiplicity detectors. 
Each hybrid carries one or more chips with the chip inputs directly wire bonded to the sensors. 
The hybrid also has a small number of passive components to adjust  the chip 
supply currents from common voltage rails and to filter the power supplies. 
The data are read out by Front End Controllers (FEC's) 
which digitize the differential analog signals in a 12 bit ADC running 
at up to 5 MHz and store the ADC output in a FIFO. 
The digitized data are passed through a G-Link connection and multiplexing boards to 
a Mercury Raceway system \cite{mercury} in which multiple PowerPC processors 
work concurrently on the data reduction. 
The FEC's also monitor chip supply voltages, detector bias voltage and leakage currents. 
The monitoring values are automaticaly transmitted together with the ADC data. 
Finally, the processed data are sent to a workstation where they can be analyzed.
\vspace*{-0.5cm}
\subsubsection{ Module Assembly and Tests}
\vspace*{-0.2cm}
After each chip is tested and accepted, it is 
manually attached with conductive glue to the hybrid. 
The chip control side is wire bonded to the hybrid. 
After bonding, the hybrid is tested by reading the pedestal, measuring the chip noise 
and running a full calibration cycle for each channel. 
Following these tests, the sensor is glued to the hybrid. Finally, the sensor is 
wire bonded to the chip input.
After assembly, the functionality of the completed module is tested in the 
same manner as the unassembled hybrid.
In addition, each module is tested with an electron source 
$^{113}$Sn (E$_{e^{-}}$ = 363 keV and 380 keV), 
which gives information about the full depletion voltage.
The signal distribution from one pad of a ring-counter sensor from a $^{113}$Sn source is presented in 
Fig. \ref{fig:signalnoise}. 
The response of a vertex sensor to minimum ionizing particles is shown in Fig. \ref{fig:cosmic}, 
which displays an energy-loss spectra for cosmic-ray muons. The peak of the minimum 
ionizing energy distribution accurs at approximately 80-90 keV. The width of the 
curve represented in the Fig. \ref{fig:cosmic} is due to Landau Fluctuations.
The signal signal-to-noise ration is approximatively 15. 
These silicon detectors have been also used during the engineering run in 1999. 
The stability of one octagon module during this period is illustrated in Fig.~\ref{fig:timepednoise}.
\vspace*{-0.5cm}
\subsection{ Performance of Silicon pad Detectors during Engineering Run at RHIC}
\vspace*{-0.5cm}
During the engineering run of June 1999 at RHIC, 
13 octagon modules, one half of the top layer of the inner vertex array (one module), 
one quarter of the top layer of the outer vertex array (two modules) 
and four first planes of spectrometer were installed. 
During this period, Au beam at injection energy (10.8 GeV/n) was circulated 
in each of the RHIC rings but not simultaneously. 
During this exposure, the silicon pad detectors
showed low noise and very good 
stability in pedestal and gain. The average pedestal and noise per 
channel for one octagon module during running 
are presented in Fig.~\ref{fig:timepednoise}. 
\newpage
\section{Summary }
\vspace*{-0.5cm}
Since the signature of the quark-gluon plasma cannot be predicted 
unabigously, PHOBOS is initially attempting to study these collisions in an unbiased way. 
Therefore PHOBOS will study the production of all types of hadronic particles. 
The majority of the emitted charged hadrons will be detected by the multiplicity detector, 
which covers almost 4$\pi$ of solid angle. This permits us to study their 
distribution in the pseudo-rapidity range of $\vert \eta \vert$ $\le$ 5.3 
on an event-by-event basis. One percent of all emitted particles will 
be studied in detail by the two-arm magnetic spectrometer in the 
mid-rapidity region, where the highest energy densities are
 expected. 
In the spectrometer, particles will be momentum analysed and identified by energy loss 
measurements in the Silicon layers. The identification of higher momentum particles 
will be augmented by time-of-flight measurement.
\\
For the silicon pad detectors, the sensor quality was demonstrated to be adequate for the 
required physics measurement. The sensor production, testing and assembly have been completed 
on schedule.
Source and beam tests show an acceptable signal-to-noise 
ratio around 15. The expected capacitive coupling between the channels 
gives a modest crosstalk of less than 1$\%$. 
\par
Currently, the silicon pad detectors of PHOBOS are ready for implementation in the RHIC 
research program. The full multiplicity detector, the vertex detectors and one arm of 
the spectrometer are ready and will be installed when RHIC physics 
running starts in June 2000.

\newpage
\vglue 3cm 

\begin{figure*}[htb]
\noindent\caption{Schematic drawing of the PHOBOS detector showing 
the major components.   
For reasons of clarity, the top half of the magnet has been removed.}
\label{fig:phobos}
\end{figure*}

\vskip 1cm

\begin{figure*}[htb]
\noindent\caption{Schematic diagram of the cross section on one pad of the PHOBOS silicon detectors.}
\label{fig:structure}
\end{figure*}

\vskip 1cm

\begin{figure*}[htb]
\caption{Schematic layout of some of the silicon sensors used in PHOBOS.\ \ \ \ \ \ \ \ \ \ \ \  }
\label{fig:octag}
\end{figure*}

\vskip 1cm

\begin{figure*}[htb]
\noindent\caption{Test results of one octagon silicon sensor. The (a), (b) and (c) plots 
represent IV test, polysilicon test and CV test, respectively.}
\label{fig:octagtest}
\end{figure*}

\vskip 1cm

\begin{figure*}[htb]
\caption{Signal distribution on one pad of the ring counter using a $^{113}$Sn source. \ \ \ \
}
\label{fig:signalnoise}
\end{figure*}

\vskip 1cm

\begin{figure*}[htb]
\hspace*{-3cm}\caption{Energy loss distribution for cosmic-ray muons on one vertex sensor. \ \ \ \ \ \ \ \ \ \  }
\label{fig:cosmic}
\end{figure*}

\vskip 1cm

\begin{figure*}[htb]
\noindent\caption{Average pedestal per channel (a) and average noise per channel 
without common mode noise subtraction (b) for one octagon module installed 
during the engineering run at RHIC.}
\label{fig:timepednoise}
\end{figure*}

\newpage
\vglue 5cm
\begin{figure*}[htb]
\psfig{figure=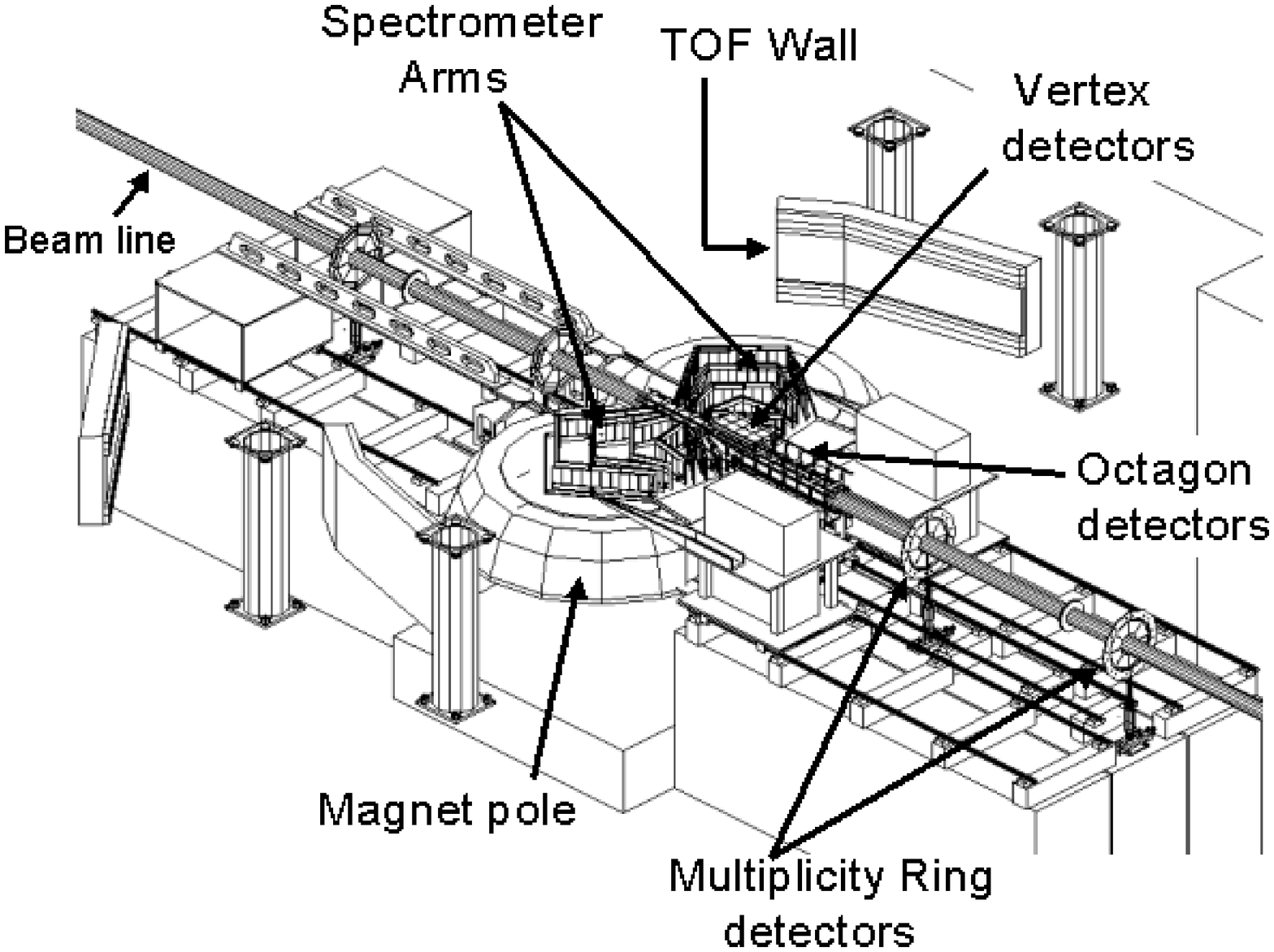,width=160mm,angle=0.0}
\end{figure*}
\vskip 3cm
\centerline{\huge \bf Fig.1}

\newpage
\vglue 5cm
\begin{figure*}[htb]
\hspace*{-0.5cm}\psfig{figure=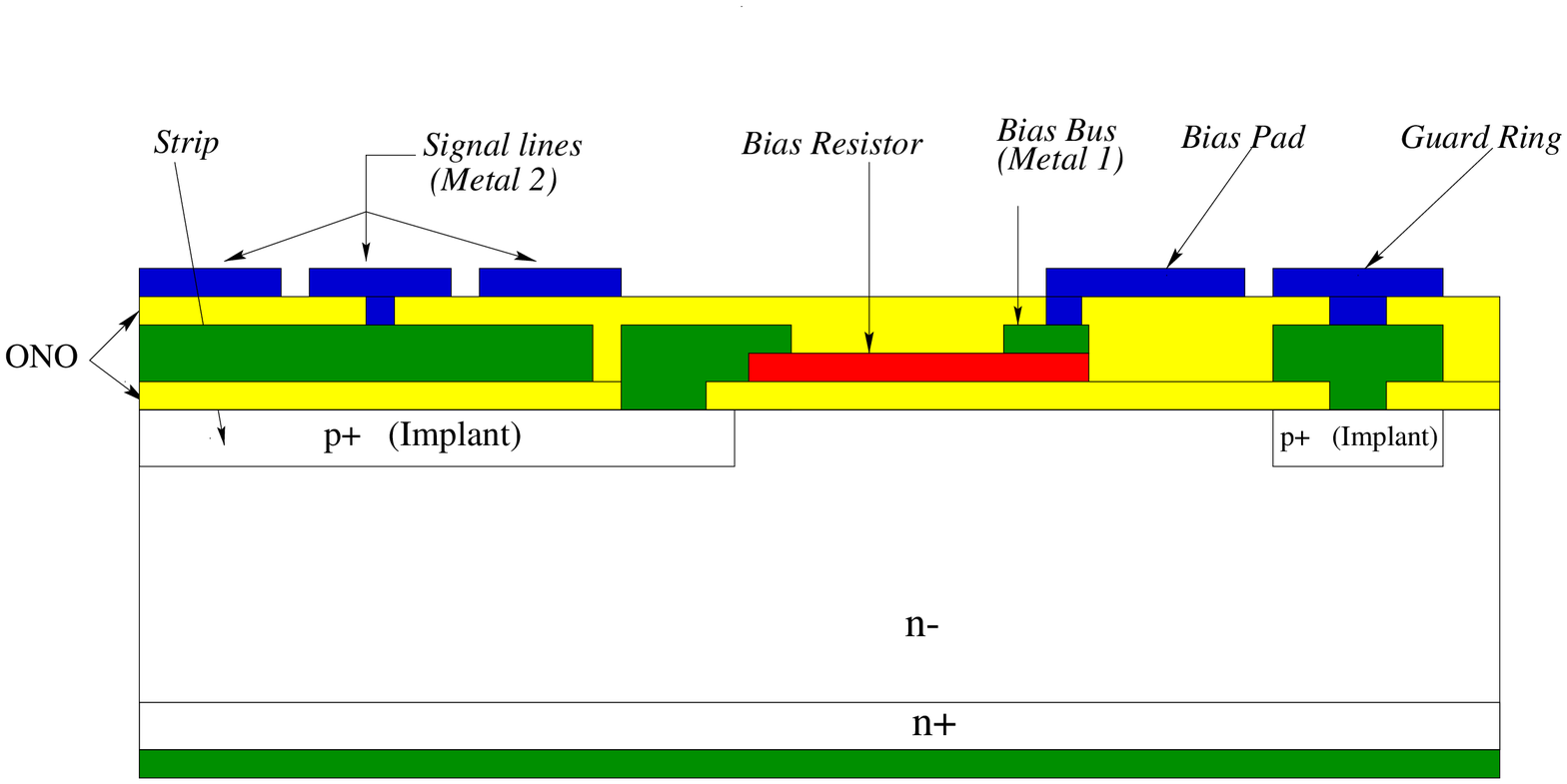,width=165mm,angle=0.0}
\end{figure*}
\vskip 3cm
\hspace*{7cm}{\huge \bf Fig.2}

\newpage
\vglue 4cm
\begin{figure*}[htb]
\psfig{figure=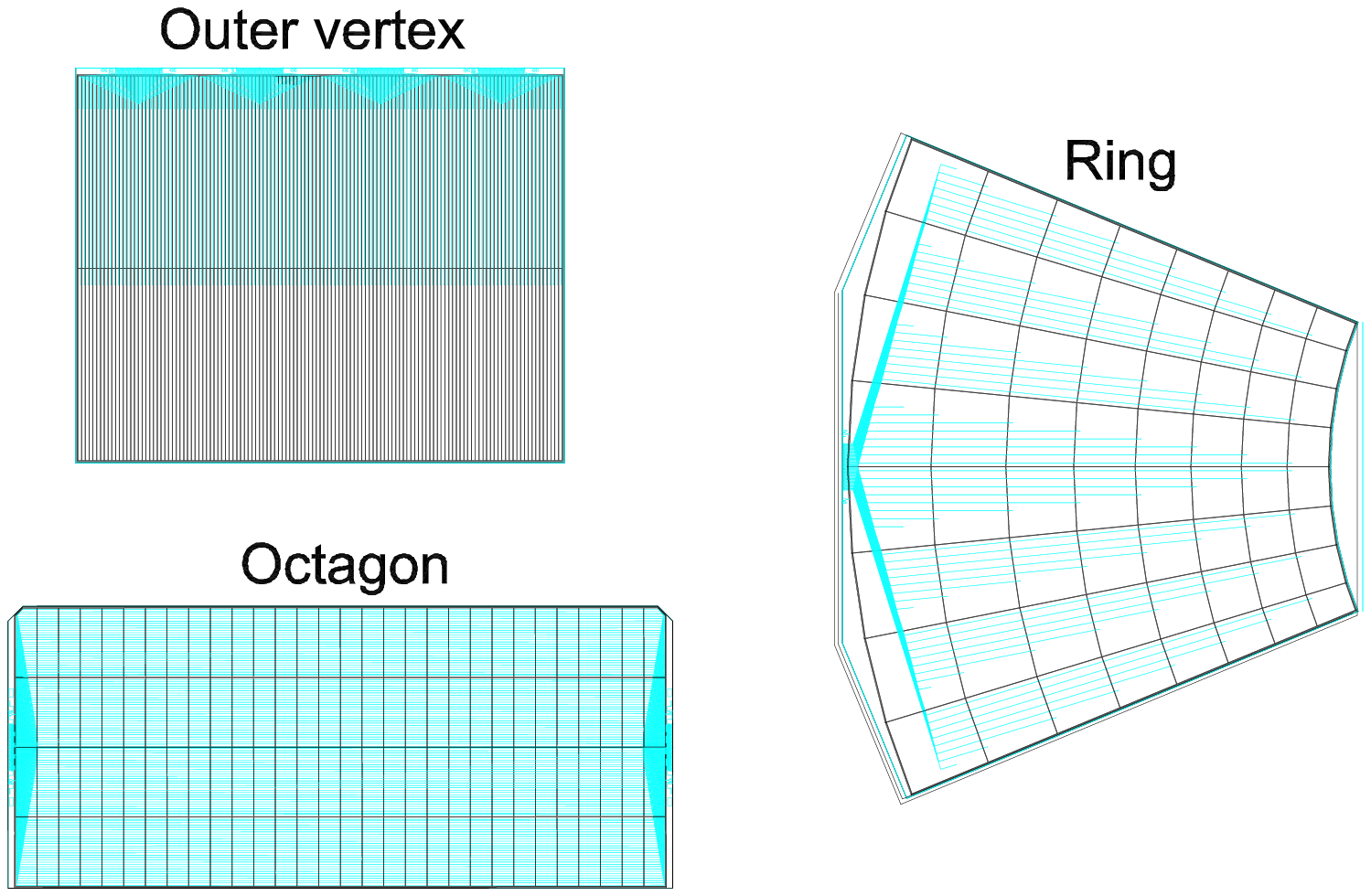,width=190mm,angle=0.0}
\end{figure*}
\vskip 3cm
\centerline{\huge \bf Fig.3}

\newpage
\vglue 4cm
\begin{figure*}[htb]
\hspace*{-0.5cm}\psfig{figure=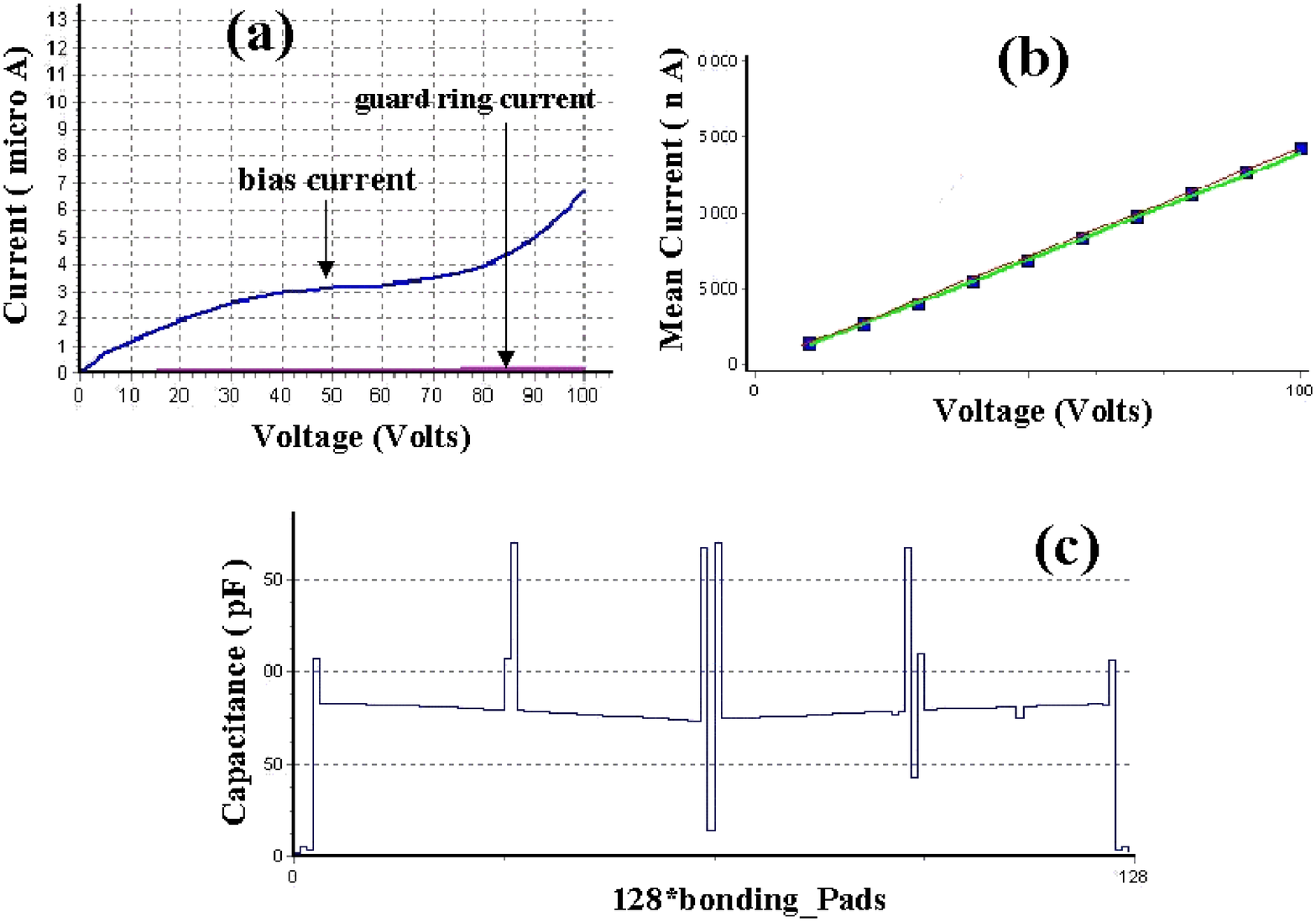,width=180mm,angle=0.0}
\end{figure*}
\vskip 3cm
\centerline{\huge \bf Fig.4}

\newpage
\vglue 3cm
\begin{figure*}[htb]
\hspace*{0.0cm}\psfig{figure=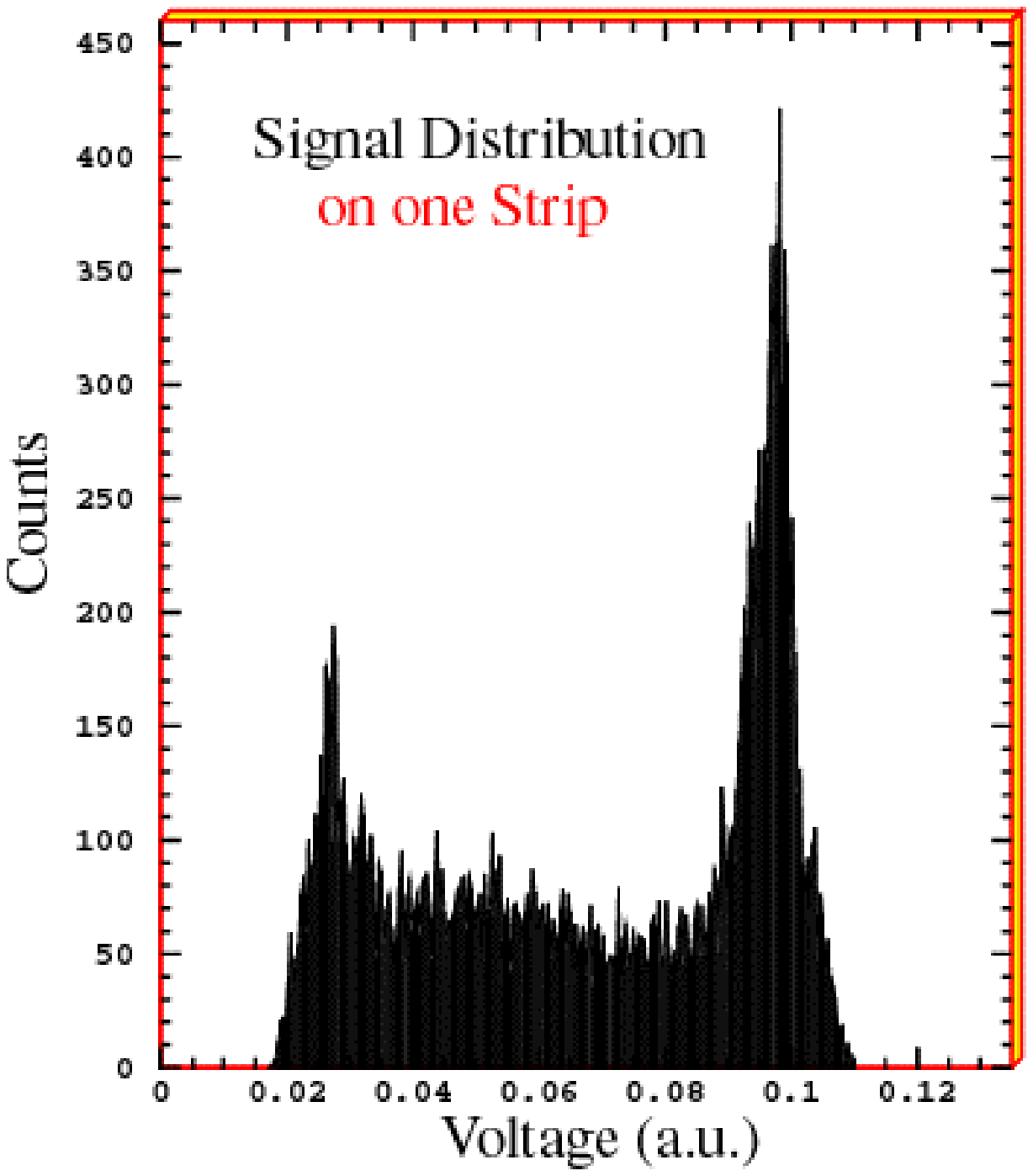,width=120mm,angle=0.0}
\end{figure*}
\vskip 3cm
\centerline{\huge \bf Fig.5}

\newpage
\vglue 3cm
\begin{figure*}[htb]
\hspace*{-2.0cm}\psfig{figure=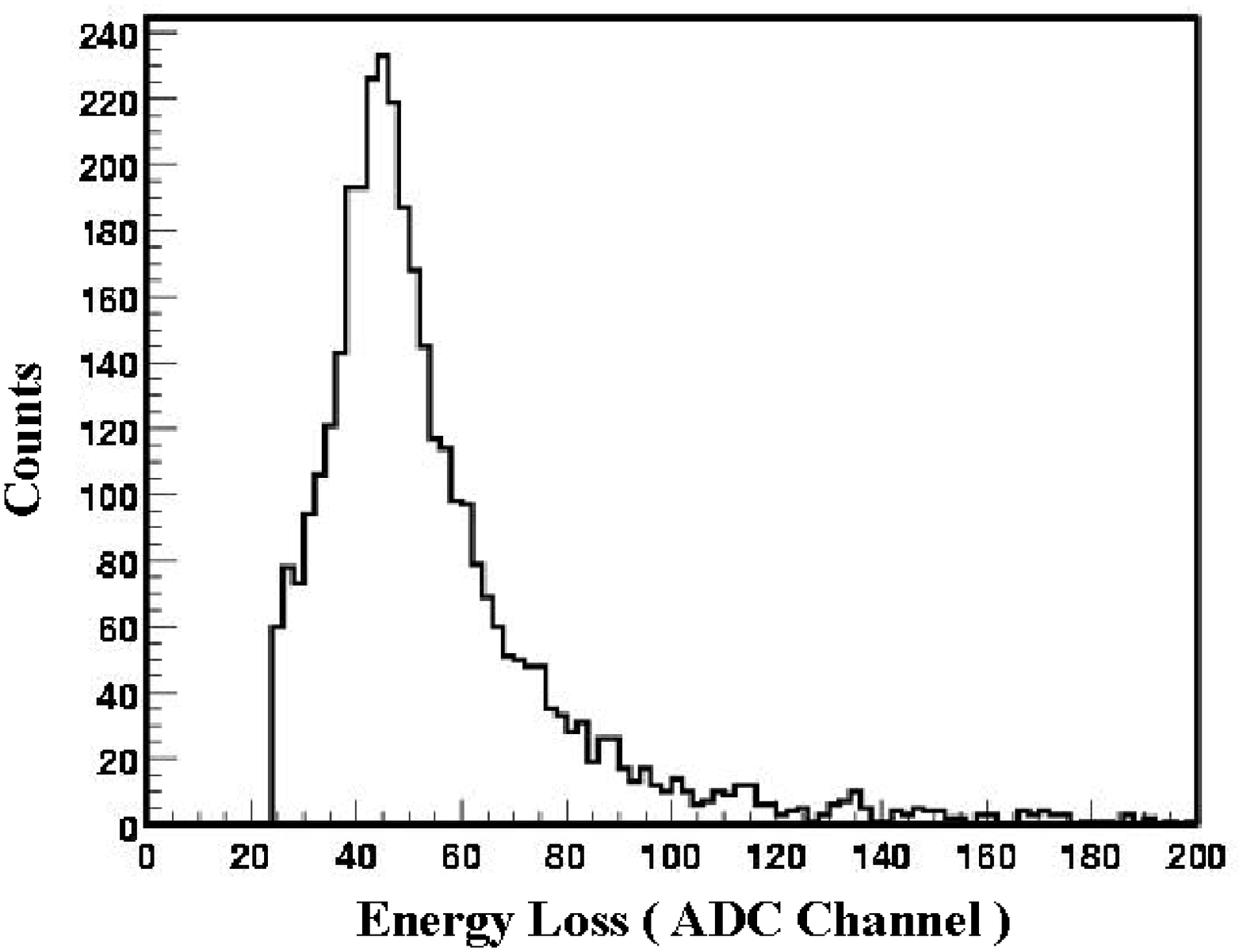,width=210mm,angle=0.0}
\end{figure*}
\vskip 1cm
\hspace*{7cm}{\huge \bf Fig.6}

\newpage
\vglue 2cm
\begin{figure*}[htb]
\hspace*{-0.5cm}\psfig{figure=fig7nouicer.eps,width=160mm,angle=-90.0}
\end{figure*}
\vskip 2cm
\hspace*{7cm}{\huge \bf Fig.7}


\vspace*{-0.4cm}
\begin{thebibliography}{999}
\vspace*{-0.3cm}
\bibitem{singh}C.P.Singh Phys. Rep. 236(1993)147.
\bibitem{harris}J. Harris and B. Muller. Ann. Rev. Nucl. Part. Sci. 46(1996)71.
\bibitem{alam}J. Alam, S. Raha and B. Sinha. Phys. Rep. 273(1996)243.
\bibitem{wilis}W.T. Lin et al., Nucl. Inst. and Meth. A 389 (1997)415.
\bibitem{gerrit}B. Back et al., Nucl. phys B78(1999)245.
\bibitem{heinz} Heinz Pernegger, Nucl. Instr. and Meth. A 419 (1998)549.
\bibitem{data} ORACLE, 500 Oracle Parkway, Redwood Shores, CA 94065, U.S.A.
\bibitem{ideas}Integrated Detector and Electronics, Veritasvein 9, N-1322 Hovik, Norway. 
\bibitem{mercury} Mercury Computer Systems, Inc., 199 Riverneck Road,
              Chelmsford, MA 01824-2820, USA.
\end{thebibliography}
\end{document}